\documentclass[runningheads]{llncs}
\usepackage[T1]{fontenc}
\usepackage{graphicx}   % graphics package
\usepackage{amsmath}    % Math support
\usepackage{hyperref}   % Hyperlinks 
\usepackage{color}

\usepackage[square,numbers]{natbib}
\usepackage[table,svgnames]{xcolor}
\usepackage{setspace}

\usepackage{subcaption}

\usepackage{booktabs} 
\usepackage{longtable} 
\usepackage{array}   

\usepackage[export]{adjustbox}

\usepackage{doi}

\usepackage{geometry}
\usepackage{listings}
\usepackage{xcolor}

\definecolor{codegreen}{rgb}{0,0.6,0}
\definecolor{codegray}{rgb}{0.5,0.5,0.5}
\definecolor{codepurple}{rgb}{0.58,0,0.82}
\definecolor{backcolour}{rgb}{0.95,0.95,0.92}
\definecolor{keywordblue}{rgb}{0.13, 0.13, 1}
\definecolor{variableorange}{rgb}{1, 0.5, 0}

\lstdefinestyle{sparqlstyle}{
    backgroundcolor=\color{backcolour},   
    commentstyle=\color{codegreen},
    keywordstyle=\color{keywordblue}\bfseries,
    stringstyle=\color{codepurple},
    numberstyle=\tiny\color{codegray},
    basicstyle=\ttfamily\scriptsize,
    breakatwhitespace=false,         
    breaklines=true,                 
    captionpos=b,                    
    keepspaces=true,                 
    numbers=left,                    
    numbersep=5pt,                  
    showspaces=false,                
    showstringspaces=false,
    showtabs=false,                  
    tabsize=2,
    morekeywords={SELECT, WHERE, PREFIX, DISTINCT, rdf, ai4se, type, hasLevel, hasTarget},
    comment=[l]{\#}, % Comments start with #
    moredelim=**[is][\color{variableorange}]{?}{ }, % Style for variables like ?paper
}

\begin{document}
\title{Navigating the growing field of research on AI for software testing -- the taxonomy for AI-augmented software testing and an ontology-driven literature survey}
\titlerunning{The \textit{ai4st} taxonomy and its use}

\author{Ina K. Schieferdecker\inst{1}\orcidID{0000-0001-6298-2327}}
\authorrunning{I. Schieferdecker}

\institute{Technische Universit{\"a}t Berlin, Einsteinufer 25, 10587 Berlin, Germany\\
\email{ina.schieferdecker@tu-berlin.de}\\
\url{https://www.tu-berlin.de}}
\maketitle    
\begin{abstract}
In industry, software testing is the primary method to verify and validate the functionality, performance, security, usability, and so on, of software-based systems. Test automation has gained increasing attention in industry over the last decade, following decades of intense research into test automation and model-based testing. However, designing, developing, maintaining and evolving test automation is a considerable effort. Meanwhile, AI's breakthroughs in many engineering fields are opening up new perspectives for software testing, for both manual and automated testing. This paper reviews recent research on AI augmentation in software test automation, from no automation to full automation. It also discusses new forms of testing made possible by AI. Based on this, the newly developed taxonomy, \textit{ai4st}, is presented and used to classify recent research and identify open research questions.
\keywords{Software Testing \and Test Automation \and Artificial Intelligence \and \textit{ai4st} \and Software Engineering \and Software Quality \and Ontology \and Semantic Web}
\end{abstract}

\section{Introduction}

\begin{flushright}
\begin{minipage}[t]{7cm}  
\footnotesize  
\begin{spacing}{1} 
"Testing with AI will take software test automation to the next level."\end{spacing}
\end{minipage}
\end{flushright}

The integration of artificial intelligence (AI) within the fields of software engineering and software testing has become a prominent area of research. Various novel testing methods and tools are being discussed and presented, yet there is a lack of comprehensive reviews addressing the range of options for augmenting software testing with AI. 
The present paper aims to address this gap by providing an introductory overview of software testing, which forms the foundation for elaborating novel testing methods enabled by AI and for developing a taxonomy on AI for software testing (\textit{ai4st}): 

Software testing is an essential part of the software development life cycle (SDLC) to ensure that software products are released with sufficient quality, reduced risks and minimized number of defects contained. Testing addresses both the verification of the software under test, i.e. whether  the software is technically of high quality, and its validation, i.e. whether the software meets the requirements. Testing uses methods of code analysis, also known as static testing because the software is not being run, and execution analysis, also known as dynamic testing because the software is being run. Dynamic testing is conducted at different levels of software composition, for example, at component level for basic software components such as classes in object-oriented programming or functions in functional programming, at integration level for compositions of software components, and at system level where the complete software-based system is tested. Software testing can be performed manually for first impressions or when test automation is not cost-effective, and automatically when testing requires automation, for example for real-time testing, or when test repetition is more efficient with test automation. Test cases, executed manually or automatically, are designed and/or generated from software requirements, software designs, bug reports or other sources of information, collectively known as the test basis. Test cases can be defined abstractly, i.e. logically at a high level with abstract preconditions, inputs, expected outputs and expected postconditions, or concretely, i.e. at a lower level with detailed preconditions, inputs, expected outputs and expected postconditions. Test cases may also contain timing requirements or other actions and procedures to make them executable. Sets of test cases form test suites. Their execution is logged and analysed for missed preconditions or mismatches with expected outputs or postconditions. The evaluation of the test runs, including the mismatches, verifies the correctness of the test cases or reports failures of the software under test. Debugging is used to identify the root causes of these failures, i.e. the errors that lead to them.

Furthermore, effective monitoring and management of the overall testing process must be implemented to ensure or enhance the quality of the software produced. In close relation to the SDLC, the testing processes can vary in terms of size, phases, teams, and so forth. They can be sequential, iterative, agile, or they can follow the principles of the W-model~\cite{spillner2002w}. However, they share the common phases of testing as defined by the fundamental testing process~\cite{spillner2021software} being planning and control, analysis and design, implementation and execution, evaluation and reporting, and completion and teardown.

However, the utilisation of AI in testing is not fully realised when it is confined solely to these phases of the fundamental test process. While it is imperative to acknowledge that testing requires also more general approaches, such as project management or technical infrastructural management, which can also be improved with AI, it is necessary to delve further into the particularities of testing. And indeed, research publications on software testing with AI is steadily growing and so is the understanding of where and how to apply AI in testing is increasing. However, an overarching view relating AI support to the various testing activities and their processes is missing. 

Taxonomies, and their formalisation as ontologies, are particularly useful for classifying and consolidating knowledge. They are powerful tools for organising, presenting and using knowledge effectively. They can act as a foundation for reasoning, interoperability and intelligent data usage. Ontologies are formalised, machine-readable representations of knowledge, extending and formalising taxonomies with additional semantics, constraints and logic. Unlike hierarchical taxonomies, ontologies enable richer querying, inference and relationship discovery.

Therefore, after reviewing taxonomies and ontologies on software testing in general and the use of AI for testing in particular in Section~\ref{Sec_Related}, an ontology dedicated to AI for software testing was developed and is presented in Section~\ref{Sec_ai4st}. Referred to as \textit{ai4st}, this ontology can be used to classify contributions from research papers on using AI techniques to evolve and/or improve software testing activities and processes. Selected results are presented and discussed  in Section~\ref{Sec_Class}. The paper concludes with an outline of future work in Section~\ref{Sec_Outlook}.

\section{Related Work}\label{Sec_Related}

According to~\cite{usman_taxonomies_2017}, taxonomies have been proposed for every knowledge area of software engineering (SE) within the SE body of knowledge. These SE knowledge areas are defined in the SWEBOK~\cite{washizaki_guide_2024} and include among others software testing fundamentals, test levels, test techniques, test-related measures, test process, software testing in the development processes and the application domains, testing of and testing through emerging technologies, and software testing tools.

However, although the highly detailed SWEBOK provides extensive descriptions of software engineering concepts, methods and techniques, including those for software testing, it does not define a comprehensive glossary or coherent taxonomy. This is somewhat surprising given the intensive discussion of the development of a SWEBOK ontology~\cite{wille_quality_2003,sicilia_evaluation_2005,abran_engineering_2006}. According to~\cite{usman_taxonomies_2017}, no other overarching ontology besides SWEBOK has been proposed, except for numerous taxonomies specific to certain knowledge areas including software testing~\cite{robinson_taxonomy_2011,utting_taxonomy_2012,felderer_taxonomy_2014,calvo_taxonomy_2015,felderer_model_2016,engstrom_serp_2017}.

Taxonomies can be formally defined using first-order logic, the Web Ontology Language OWL~\cite{mcguinness_owl_2004} or the UML-based Ontology Modelling Language OntoUML~\cite{guizzardi_endurant_2018}. OWL has attracted considerable interest, particularly due to its association with the Semantic Web and the support offered by the Protégé tool. Nevertheless, in most cases, there is no formally defined ontology available for taxonomies proposed in the literature that can be reused for classification purposes, e.g. for the classification of software test research contributions. 

In addition, although numerous ontologies related to software testing (ST) exist~\cite{tebes_analyzing_2020}, they are limited in their coverage of the research field, their ability to classify research contributions, and/or their relation to established glossaries and/or taxonomies: According to~\cite{calero_ontologies_2006}, ontologies for SE including ST are either generic, such as the software engineering ontology network SEON~\cite{borges_seon_2016}, or specific to a knowledge area, such as software testing like the Reference Ontology on Software Testing ROoST~\cite{souza_roost_2017}. 

It is important to note that these two ontologies (and others, such as OntoTest~\cite{barbosa_towards_2006} and TestTDO~\cite{tebes_testtdo_2020}) focus on the conceptual grounding of SE or ST concepts, respectively, with regard to their philosophical relations, rather than than focusing on the established body of knowledge. A body of knowledge provides not just terms and relations, but also definitions, explanations, examples and/or best practices. Relevant bodies of knowledge for software testing include the SWEBOK~\cite{washizaki_guide_2024}, the SE terms by ISO, IEC and IEEE in~\cite{iso_24765standard_2017}, also known as SEVOCAB, and the ST terms by ISTQB in~\cite{istqb_glossary_2025}, also known as the ISTQB Glossary. 

Furthermore, ROoST~\cite{souza_roost_2017}, being also part of SEON, OntoTest~\cite{barbosa_towards_2006} or the ontology presented in~\cite{zhu_developing_2005} focus on dynamic testing only. Although TestTDO~\cite{tebes_testtdo_2020} includes static testing, it does not link to bodies of knowledge either. 

It should also be noted that large-scale research taxonomies such as the Computer Science Ontology CSO~\cite{salatino_computer_2020} do not detail software testing. It categorises 'Testing and Debugging' as a research topic comprising  15 subtopics only. Similarly, the ACM Computing Classification System~\cite{rous_major_2012} only covers some aspects of software testing research.

In summary, and to the best of the author's knowledge, there is no software testing ontology that
\begin{itemize}
	\item is formally defined, machine-processable, and downloadable as OWL (or in a comparable format),  
	\item can be reused and extended for new application scenarios like software testing research classification, 
	\item is closely linked to well-established and standardized bodies of knowledge, and 
	\item covers the software testing knowledge area extensively.
\end{itemize}

The development of this ontology is described in the following section.

\section{The Ontology for Artificial Intelligence in Software Testing -- \textit{ai4st}}\label{Sec_ai4st}

In order to support the classification of research contributions in the area of using AI methods and techniques for the improvement of ST activities and processes, a dedicated taxonomy named \textit{ai4st} has been developed by

\begin{itemize}
	\item making primarily use of the terms defined by ISTQB~\cite{istqb_glossary_2025}\footnote{Where necessary, these terms are supplemented by SEVOCAB terms (see ISO 24765:2017). Where research exceeds the scope of standardised terminology, the author has provided the most accurate description possible.}, 
	\item defining it in OWL\cite{mcguinness_owl_2004} to support machine-processability,
	\item assigning the CC-BY-SA license~\cite{cc_2025} to support reuse, and
	\item providing it via GitHub~\cite{schieferdecker_ai4st_2025} to enable contributions and/or the uptake by others.
\end{itemize}

\noindent Furthermore, the \textit{ai4st} taxonomy is based on a four layer model consisting of 
\begin{itemize}
	\item the lightweight universal foundational ontology gUFO~\cite{almeida_gufo_2019} for grounding,
	\item the software testing concept ontology stc~\cite{schieferdecker_stc_2025},
	\item the AI for software engineering ontology \textit{ai4se}~\cite{schieferdecker_next_2024}, and
	\item the consolidated overarching \textit{ai4st} ontology itself.
\end{itemize}

The ontology gUFO~\cite{almeida_gufo_2019} is a simplified version of the Unified Foundational Ontology UFO, designed for easier integration with ontology-driven conceptual modelling, particularly in domains such as information systems. It supports objects (enduring entities) and qualities related to them, which are used to classify research papers into the dimensions of the \textit{ai4st} taxonomy. 

The \textit{stc} ontology~\cite{schieferdecker_stc_2025} represents a selection of terms in the ISTQB glossary. The glossary contains keywords from software testing syllabi, covering foundation, advanced, and expert-level concepts, methods, and techniques in the software testing profession. Currently under development using the Protégé tool, the \textit{stc} ontology consists of over 200 classes representing software testing terms and 20 object properties representing their relations. Each term is described as being defined by ISTQB, SEVOCAB, or as proprietary. The development of \textit{stc}  began with the concept maps provided by ISTQB and has evolved beyond them, as these concept maps are informal and mainly represent top-level terms in software testing.

As a predecessor to \textit{ai4st}, the \textit{ai4se} taxonomy~\cite{schieferdecker_next_2024} was created to structure the emerging research field of applying AI to SE and to address its nuances. \textit{ai4se} is structured along four dimensions:

\begin{itemize}
	\item Purpose: The goal of using AI is to understand, generate or improve SE artefacts/processes. A new approach may address one, two, or all three of these purposes; for example, it may address both understanding and generation.
	\item Target: The SE activity is addressed in (1) development and (2) operations, as well as in the corresponding (3) processes. Whenever models are central to an approach, as they are in Model-Driven SE, this is denoted as well. An approach may target several SE activities. SE processes consist of SE activities and constitute activities themselves, allowing for a more detailed representation of an SE target.
	\item AI Type: The AI techniques being used by an approach, including Symbolic, Subsymbolic, Generative, Agentic, and General AI. One approach may use several types of AI.
	\item Level: The degree of automation is based on a five-level scale ranging from (1) no support to (5) full automation. The highest level achieved by an approach is indicated. Currently, level 3 (AI-assisted selection) is the most common.
\end{itemize}

\begin{figure}[!ht]
    \centering % This centers the entire figure block

    % --- TOP ROW ---
    \begin{subfigure}{0.48\textwidth}
        \centering
        % The key is to set the width relative to the subfigure's container
        \includegraphics[width=\linewidth, 
        max height=0.22\textheight]{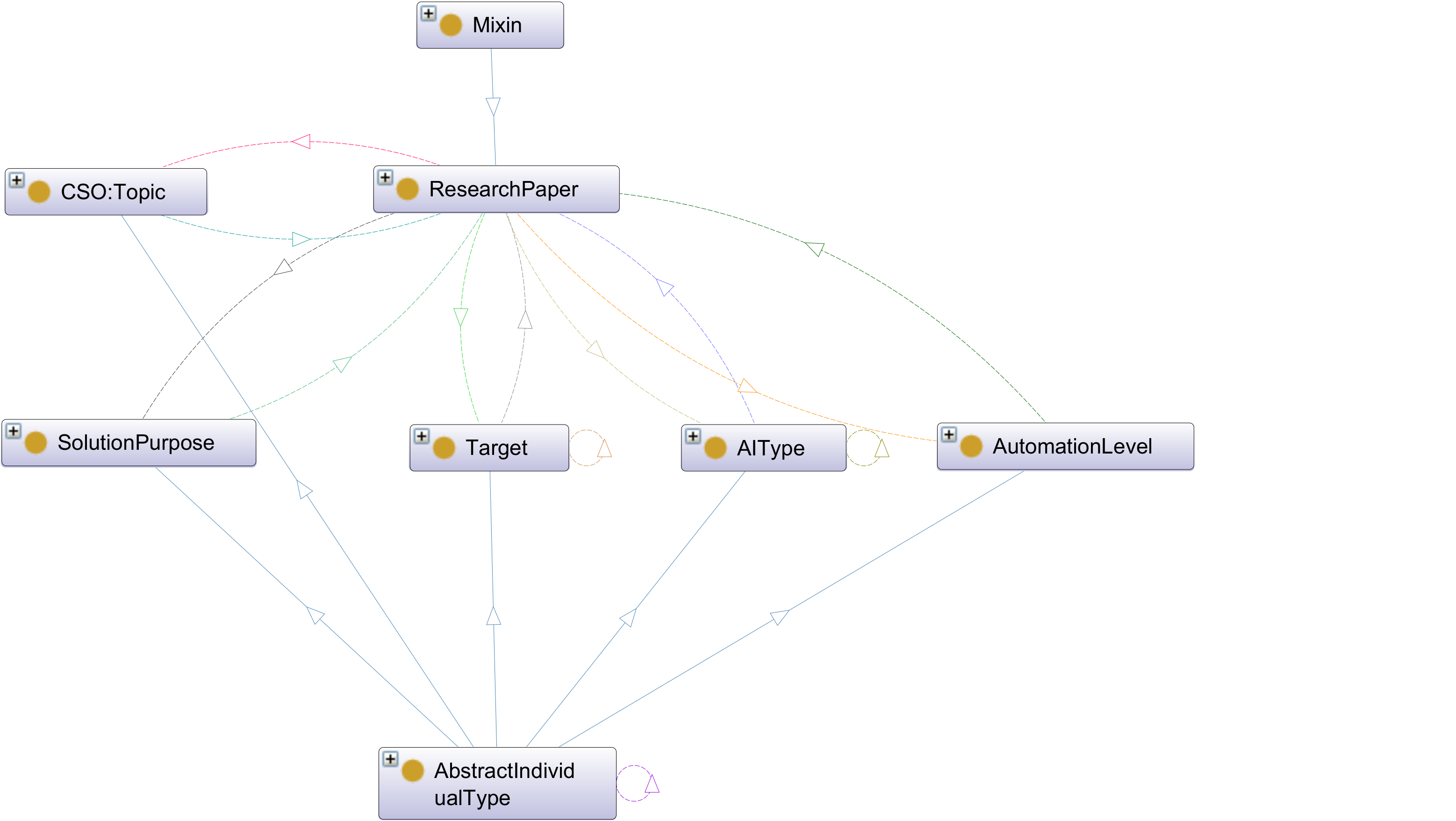} 
        \caption{The dimensions of the \textit{ai4st} taxonomy: Research Topic, Solution Purpose, ST Target, AI Type, Automation Level.}
        \label{fig:AI4STOverview}
    \end{subfigure}% <--- This % sign is important, it prevents a horizontal space
    \hfill % This adds a flexible space between the two top figures
    \begin{subfigure}{0.48\textwidth}
        \centering
        \includegraphics[
        max width=\linewidth, 
        max height=0.2\textheight
    ]{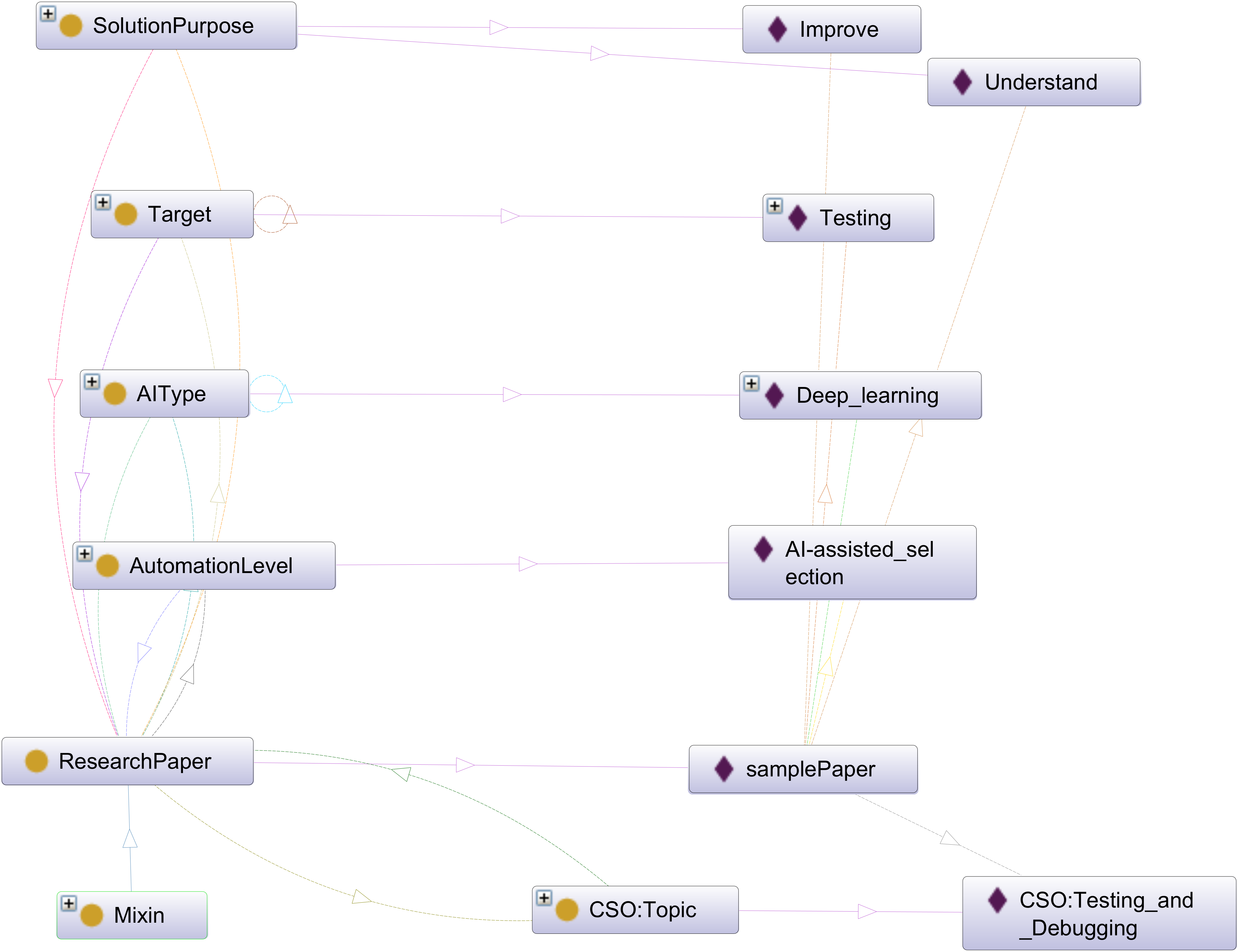}
        \caption{The classification of a sample paper about testing and debugging, aiming at understanding and improving the testing activities, using deep learning methods, and providing AI-assisted selections.}
        \label{fig:AI4STSample}
    \end{subfigure}

    % --- MIDDLE ROW ---
    \begin{subfigure}{0.48\textwidth}
        \centering
        \includegraphics[width=\linewidth, 
        max height=0.1\textheight]{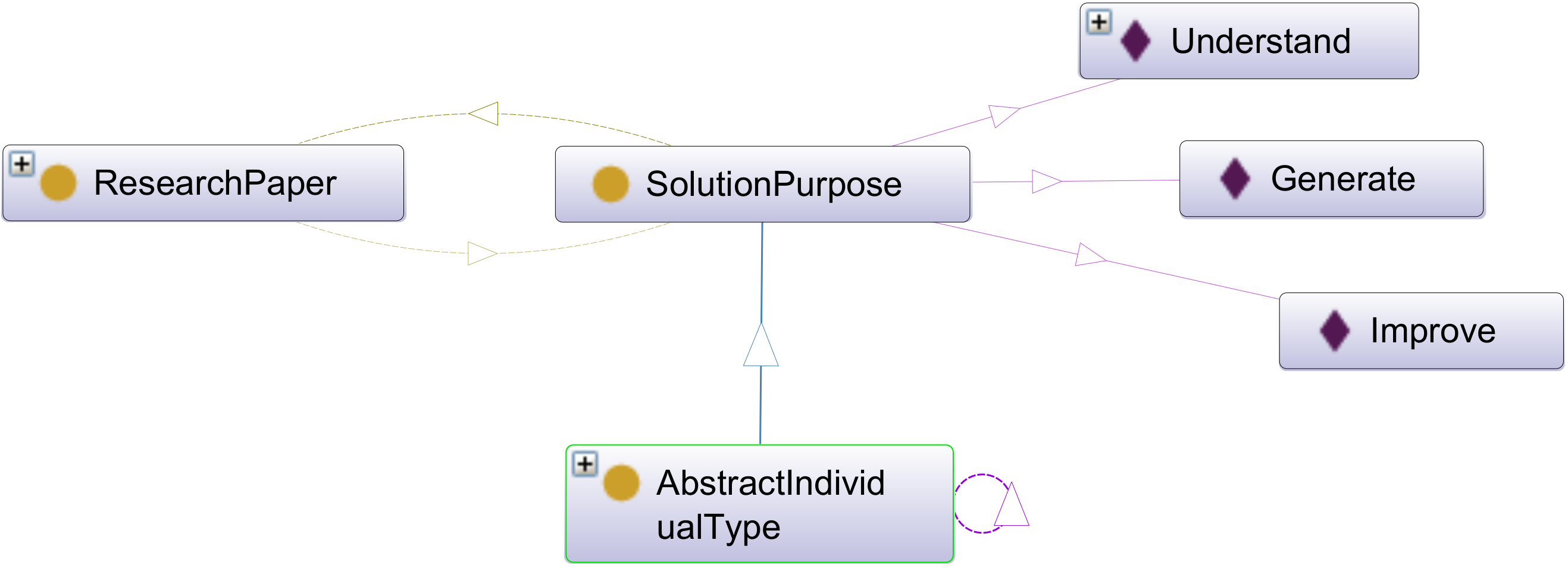}
        \caption{The solution purpose classifiers to understand, generate, or improve ST artefacts}
        \label{fig:AI4STPurpose}
    \end{subfigure}%
    \hfill
    \begin{subfigure}{0.48\textwidth}
        \centering
        \includegraphics[width=\linewidth]{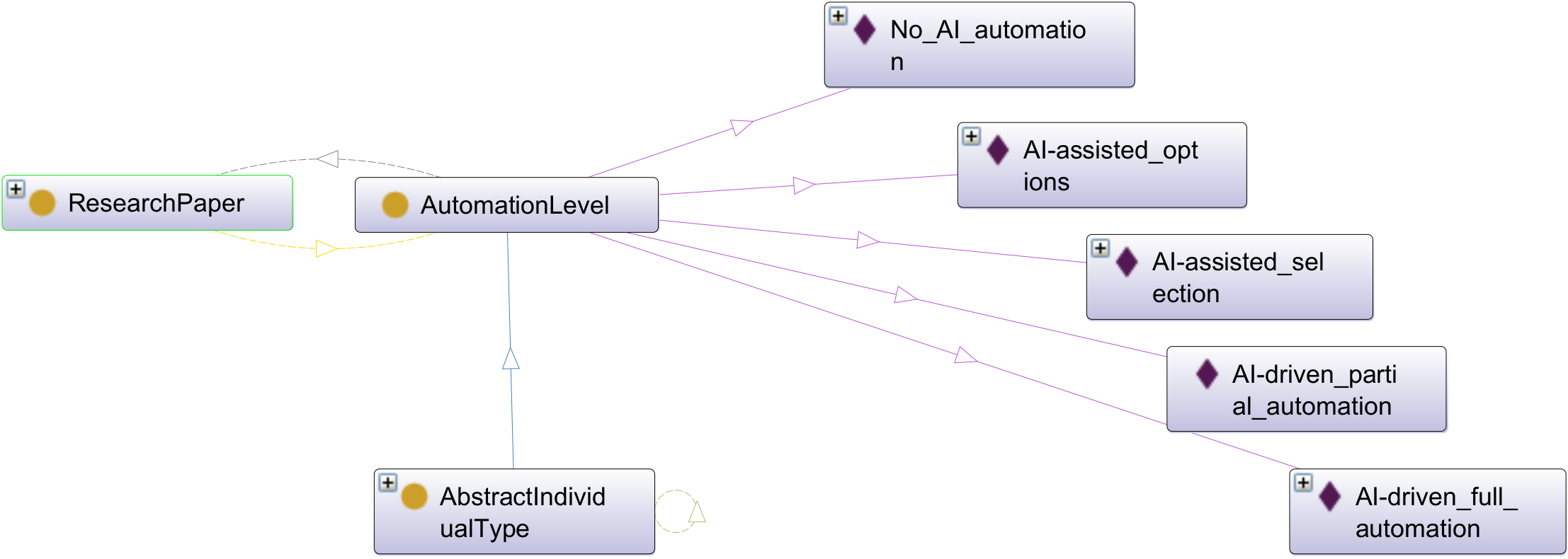}
        \caption{The automation level classifiers consisting of no AI support, providing AI-assisted options or AI-assisted selections, or supporting AI-driven partial automation or AI-driven full automation.}
        \label{fig:AI4STLevel}
    \end{subfigure}

    % --- BOTTOM ROW ---
    \begin{subfigure}{0.48\textwidth}
        \centering
        \includegraphics[
        max width=\linewidth, 
        max height=0.3\textheight
    ]{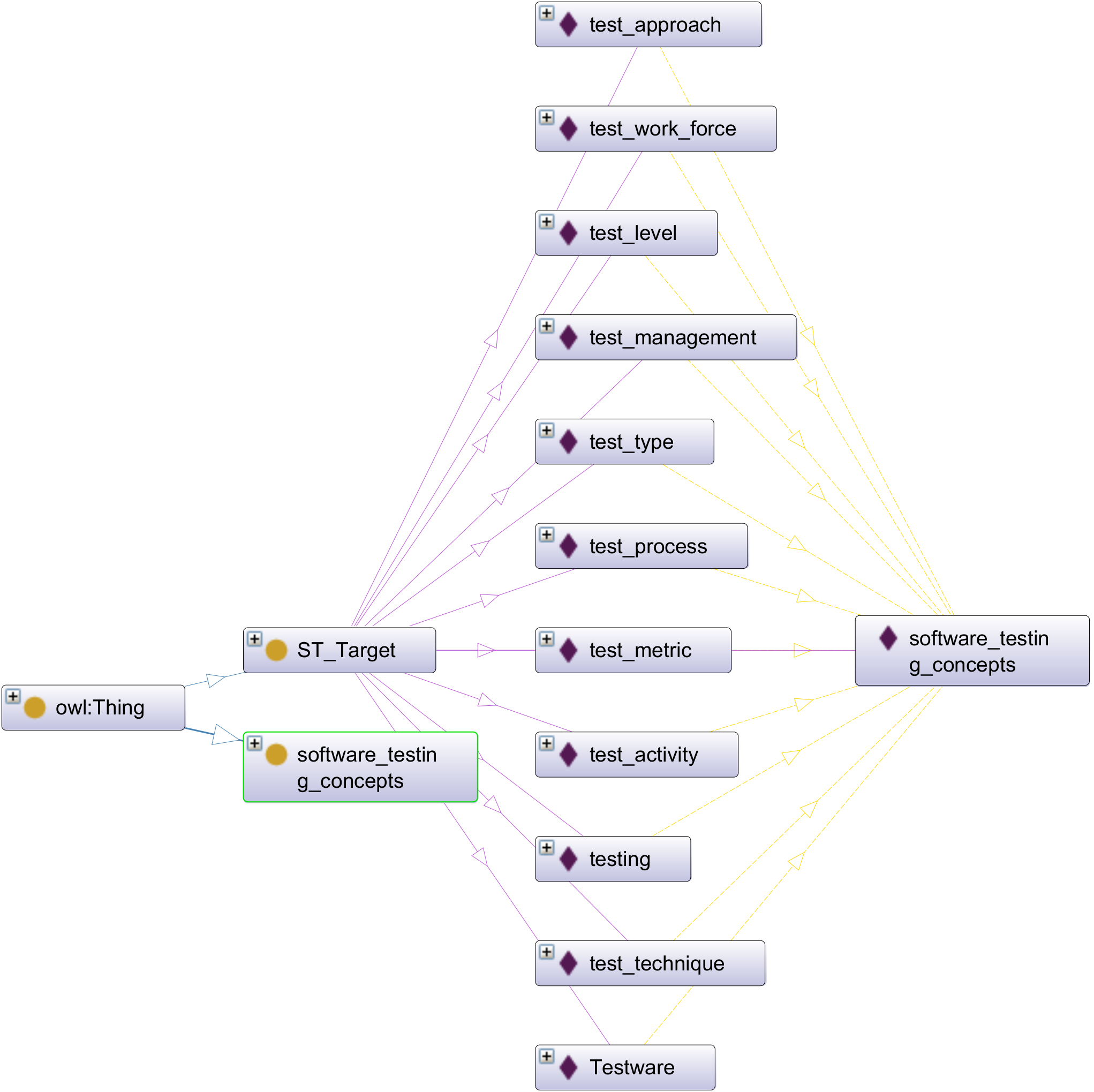}
        \caption{The target classifiers represent the ST techniques, activities, or processes to which an AI system is being applied\footnotemark}
        \label{fig:AI4STTarget}
    \end{subfigure}
		\hfill
    \begin{subfigure}{0.48\textwidth}
        \centering
        \includegraphics[width=\linewidth]{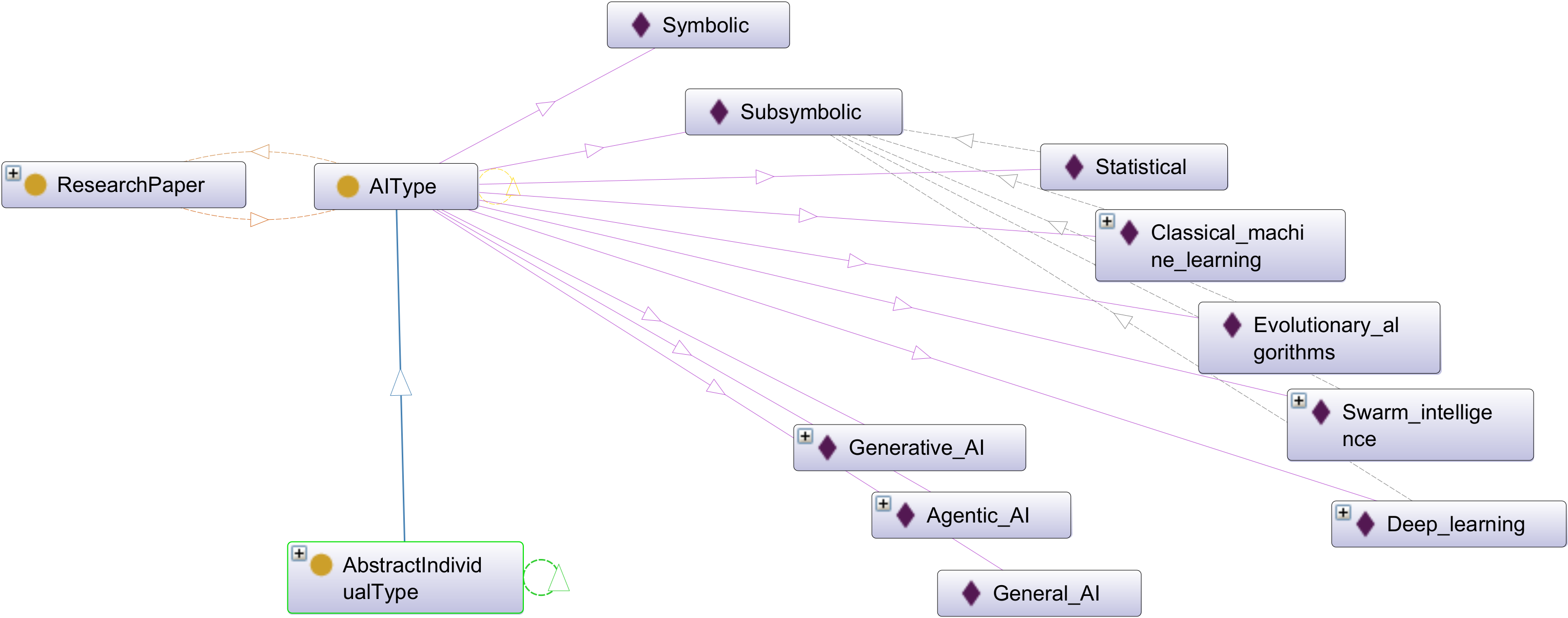}
        \caption{The AI type classifiers for sumbolic, subsymbolic, generative, agentic, or general AI. Subsymbolic AI is classified into statistical, classical machine learning, evolutionary algorithms, swarm intelligence, and deep learning AI.}
        \label{fig:AI4STType}
		\end{subfigure}

    \caption{Research Paper Classification with \textit{ai4st}} \label{figClassifying}
\end{figure}

When the \textit{ai4se} taxonomy was used to classify a large set of research papers, it became clear that a tool-based approach, such as that offered by the Semantic Web, was preferable to manual classification. Hence, the \textit{ai4se} ontology was developed~\cite{schieferdecker_ai4se_2025}. It also became apparent that, in order to address the specifics of using AI for software testing, a more specific taxonomy and machine-processable ontology for software testing research would be preferable. The resulting \textit{ai4st} ontology~\cite{schieferdecker_ai4st_2025} is shown in Figure \ref{figClassifying} with its (top-level) dimensions.

\footnotetext{For the \textit{ai4st} classification, the classes of the software testing ontology \textit{stc} have been reified/punned to individuals to represent the software testing terms. The hierarchical relationship between the concepts has been kept in a separate object property 'has parent'. These terms constitute the target classifiers in \textit{ai4st}.}

\section{Classification of AI for Software Testing Research}\label{Sec_Class}

The classification of AI for software testing research is an ongoing project. To check the validity of the \textit{ai4st} ontology, an adapted, lightweight systematic literature review (SLR), as described in~\cite{stapic_performing_2012}, was conducted to analyse related research. This SLR protocol was followed:

\begin{itemize}
	\item \textbf{Review title}: Initial SLR on the application of the \textit{ai4st} taxonomy.
	\item \textbf{Objectives of the review}: 
		\begin{enumerate}
			\item To test the validity of the \textit{ai4st} taxonomy with an initial research selection.
			\item To determine a classification of this initial research selection.
		\end{enumerate}
	\item \textbf{Research questions}:
		\begin{itemize}
			\item RQ1: Which standardized terms are being used in \textit{ai4st} related research?
			\item RQ2: Which alternative terms are being used in \textit{ai4st} related research?
			\item RQ3: Are the \textit{ai4st} taxonomy dimensions useful to classify the pre-selected research?
		\end{itemize}
	\item \textbf{Database}: Research papers from the conference proceedings of the latest International Conference on Software Engineering, ICSE 2025 in Ottawa, Canada and its co-located conferences and workshops; and referenced papers for backward snowballing the research. Forward snowballing was in this case unnecessary, as ICSE 2025 represented the most recent research publications at that time. A complementary search of the IEEE and ACM digital libraries has added further software testing and AI-related research papers published between 2020 and 2025.
	\item \textbf{Inclusion criteria}: 
		\begin{itemize}
			\item Peer-reviewed original research.
			\item Online available.
			\item Research on AI for ST.
		\end{itemize}
	\item \textbf{Exclusion criteria}: 
		\begin{itemize}
			\item Meta-research such as evaluations, benchmarking, comparisons, surveys, taxonomies, roadmaps
			\item Testing of software-based systems like IoT, cloud, vehicle, etc.
			\item Research on ST for AI.
			\item Posters and tutorials.
		\end{itemize}
	\item \textbf{Selection process}:
		\begin{enumerate}
			\item Title and abstract screening for the pre-selection of unique research candidates by use of 
				\begin{enumerate}
					\item the concept map resulting from the \textit{stc} ontology~\cite{schieferdecker_stc_2025} to identify ST-related research, and
					\item the concept map resulting from the \textit{ai4st} dimensions 'AI type' to identify AI related research in the ST-related subset of research.
				\end{enumerate}
			\item Full text review and assessment of the research contributions for the final selection of unique research.
			\item Tools: 
			\begin{itemize}
				\item Online research libraries, including dblp, ACM DL, IEEE Xplore, and Google Scholar to identify related work; and
				\item Python for text analysis and post-processing of finally selected research, supported by MS Visual Studio, Google AI Studio, and LibreOffice.
			\end{itemize}
		\end{enumerate}
	\item \textbf{Synthesis process}: 
	\begin{itemize}
		\item Review of the new and synonym candidate terms for inclusion into the \textit{stc} ontology.
		\item Classification of the selected research for inclusion into the \textit{ai4st} ontology.
	\end{itemize}
\end{itemize}

\begin{center}
	\begin{figure}[!ht]%
		\includegraphics[height=0.3\textheight]{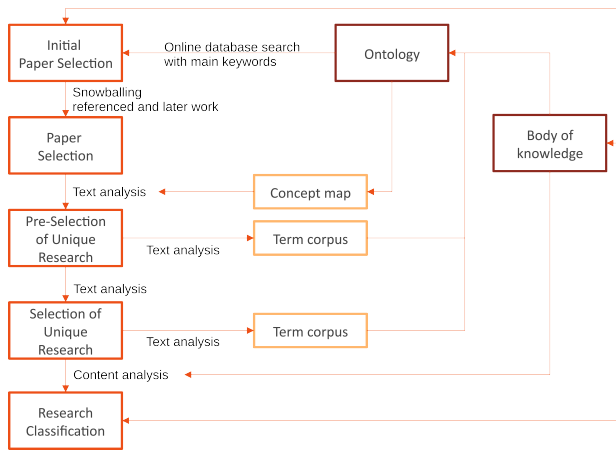}%
		\caption{Overview on the ontology-driven systematic literature review (SLR) combined with the SLR-driven ontology development}%
		\label{figOntologyDevelopment}%
	\end{figure}
\end{center}

Alongside the analysis of recent AI for ST research, a new SLR approach has been developed. Rather than using simple search expressions, the pre-selection of papers uses more detailed concept maps that are derived from the relevant ontologies and also include synonyms, as shown in Figure~\ref{figOntologyDevelopment}. Additionally, the SLR results are used to verify, improve and extend the ontologies further. Therefore, assessing the research texts also involves searching for new term and new synonym candidates. The potential for further refining this kind of ontology-based SLR, as well as SLR-based ontology refinement, depends on the features of digital library APIs that enable more powerful automated searches and research (meta-)data collections.

In result of the SLR and text (title and abstract) analysis, of the 1643 papers identified, 1150 referred to a term in the \textit{stc} ontology in their abstracts and/or titles, but 735 of these referred to only one term, indicating that the paper merely references a fact about software testing. Another 1337 papers used variations of the terms in the \textit{stc} ontology, such as 'unit test' instead of 'unit-level test'\footnote{In a future version of the \textit{stc} ontology, more synonym terms will be added to better cope with variations and synonyms of terms.}. 460 papers referred to only one variation. Papers using two or more original or alternative terms, what makes them candidate papers for the \textit{ai4st} taxonomy, form a body of 949 papers. Of these 949 papers, 656 contained terms from the \textit{ai4st} ontology related to AI. Within the 656 papers, 38 relevant original research papers for \textit{ai4st} were identified. 

Furthermore, the text analysis revealed 40 new \textit{stc} term candidates, 53 new \textit{stc} synonym candidates and 26 terms that can be treated as either new term or synonym candidates. These new term candidates include terms such as 'test result' and 'fuzz testing', which are included in the ISTQB Glossary -- a collection of over 600 terms -- but are not yet included in the initial \textit{stc} ontology, which contains over 200 terms. The new term candidates also include terms such as 'flaky test' and 'genetic testing', which are not included in the ISTQB Glossary but are extensively discussed in research. Another four new term candidates, such as 'test bot' and 'bias testing', and one new synonym candidate, 'AI-based', have been identified for the \textit{ai4st}  ontology. This is mainly because the SRL focused on ST rather than AI.  In the next release of \textit{ai4st}, the decision will be made as to  which new or synonym candidates, beyond those required for research classification in this paper (see below), will be added and whether distinguishing AI types in more detail would be useful.

\begin{footnotesize}
\begin{longtable}{m{0.05\textwidth} m{0.92\textwidth}}
\caption{List of Research Papers}\label{tab:SLR}
\label{tab:paperslist} \\
    \hline
    \textbf{Ref} & \textbf{Title} \\
    \hline
\endfirsthead
    \hline
    \multicolumn{2}{r}{\textit{...continued from previous page}} \\
    \hline
\endhead
    \hline
\endlastfoot
    \cite{de_santiago_junior_method_2022}&A Method and Experiment to evaluate Deep Neural Networks as Test Oracles for Scientific Software\\
    \cite{naimi_new_2024}& A new approach for automatic test case generation from use case diagram using LLMs and prompt engineering\\
    \cite{ferreira_acceptance_2025} & Acceptance Test Generation with Large Language Models: An Industrial Case Study \\
    \cite{de_almeida_ai_2024}&AI in Service of Software Quality: How ChatGPT and Personas Are Transforming Exploratory Testing\\
    \cite{kapoor_ai-assisted_2025}&AI-Assisted Test Script Generation for GUI Applications\\
    \cite{mohacsi_ai-based_2021}&AI-Based Enhancement of Test Models in an Industrial Model-Based Testing Tool\\
    \cite{maia_ai-driven_2024}&AI-Driven Acceptance Testing: first insights exploring the educational potential for test analysts\\
    \cite{helmy_ai-driven_2024}&AI-Driven Testing: Unleashing Autonomous Systems for Superior Software Quality Using Generative AI\\
    \cite{martin-lopez_ai-driven_2020}&AI-driven web API testing\\
    \cite{garlapati_ai-powered_2024}&AI-Powered Multi-Agent Framework for Automated Unit Test Case Generation: Enhancing Software Quality through LLMs\\
    \cite{hagar_ais_2025}&AIs Understanding of Software Test Architecture\\
    \cite{prasetya_agent-based_2021}&An Agent-based Architecture for AI-Enhanced Automated Testing for XR Systems\\
    \cite{kaur_approach_2020}&An Approach To Extract Optimal Test Cases Using AI\\
    \cite{gao_approach_2022}&An Approach to GUI Test Scenario Generation Using Machine Learning \\
    \cite{primbs_assert5_2025} & AsserT5: Test Assertion Generation Using a Fine-Tuned Code Language Model\\
    \cite{olmez_automation_2024}&Automation of Test Skeletons Within Test-Driven Development Projects\\
    \cite{yao_bugblitz-ai_2024}&BugBlitz-AI: An Intelligent QA Assistant\\
    \cite{gao_clozemaster_2025}& ClozeMaster: Fuzzing Rust Compiler by Harnessing LLMs for Infilling Masked Real Programs\\
    \cite{wang_deep_2024}&Deep Multiple Assertions Generation\\
    \cite{caglar_development_2023}&Development of Cloud and Artificial Intelligence based Software Testing Platform (ChArIoT)\\
    \cite{garg_generative_2023}&Generative AI for Software Test Modelling with a focus on ERP Software\\
    \cite{strandberg_ethical_2021}&Ethical AI-Powered Regression Test Selection\\
    \cite{happe_getting_2023}&Getting pwn.d by AI: Penetration Testing with Large Language Models\\
    \cite{zimmermann_gui-based_2023}&GUI-Based Software Testing: An Automated Approach Using GPT-4 and Selenium WebDriver\\
    \cite{calvano_leveraging_2025}&Leveraging Large Language Models for Usability Testing: a Preliminary Study\\
    \cite{franzosi_llm-based_2025}&LLM-Based Labelling of Recorded Automated GUI-Based Test Cases\\
    \cite{zhang_new_2024}&New Approaches to Automated Software Testing Based on Artificial Intelligence\\
    \cite{peixoto_effectiveness_2025}&On the Effectiveness of LLMs for Manual Test Verifications\\
    \cite{gamal_owl_2023}&Owl Eye: An AI-Driven Visual Testing Tool\\
    \cite{leu_reducing_2024}&Reducing Workload in Using AI-based API REST Test Generation\\
    \cite{confido_reinforcing_2022}&Reinforcing Penetration Testing Using AI\\
    \cite{ghimis_river_2020}&RIVER 2.0: an open-source testing framework using AI techniques\\
    \cite{abdelkarim_tcp-net_2023}&TCP-Net++: Test Case Prioritization Using End-to-End Deep Neural Networks Deployment Analysis and Enhancements\\
    \cite{hossain_togll_2024} & TOGLL: Correct and Strong Test Oracle Generation with LLMs\\
    \cite{shirzadehhajimahmood_using_2021}&Using an agent-based approach for robust automated testing of computer games\\
    \cite{de_santiago_junior_method_2022} & Using Large Language Models to Generate Concise and Understandable Test Case Summaries\\
    \cite{ragel_visual_2023}& Visual Test Framework: Enhancing Software Test Automation with Visual Artificial Intelligence and Behavioral Driven Development\\
    \cite{haldar_wip_2024}& WIP: Assessing the Effectiveness of ChatGPT in Preparatory Testing Activities\\
\end{longtable}
\end{footnotesize}

Due to space limitations, the classification of the selected 38~papers given in Table~\ref{tab:SLR} is not fully shown here. The complete results on pre-selected research and finally selected unique research, as well as the usage of terms and the assessment of new term and synonym candidates, are provided in~\cite{schieferdecker_annex_2025}. 

The research questions RQ1, RQ2, and RQ3 of this SLR are answered briefly as follows: Classifying the unique research led to an extension of the \textit{stc} ontology with:
\begin{itemize}
	\item three new terms defined in the ISTQB Glossary: 'visual testing'~\cite{gamal_owl_2023,ragel_visual_2023,prasetya_agent-based_2021}, 'assertion'~\cite{wang_deep_2024,primbs_assert5_2025}, and 'penetration testing'~\cite{happe_getting_2023,confido_reinforcing_2022}
	\item eight new terms not in the ISTQB Glossary: 
	\begin{itemize}
		\item two test techniques: 'mutation testing'~\cite{caglar_development_2023} and 'concolic testing'~\cite{ghimis_river_2020};
		\item five test activities: 'test selection'~\cite{strandberg_ethical_2021}, 'test generation'~\cite{zhang_new_2024,garg_generative_2023,gao_clozemaster_2025}, 'test verification'~\cite{peixoto_effectiveness_2025}, 'test prioritization'~\cite{abdelkarim_tcp-net_2023}, and 'test documentation'~\cite{djajadi_using_nodate};
		\item one non-functional testing: 'penetration testing'~\cite{happe_getting_2023,confido_reinforcing_2022}; and
		\item one basic concept: 'ethics'~\cite{strandberg_ethical_2021}
	\end{itemize} 
\end{itemize}

Furthermore, the new synonym 'test architecture'~\cite{hagar_ais_2025} for 'test approach' was added. In response to RQ1 and RQ2, the software testing targets in the unique research papers were successfully classified by combining the new terms and synonyms with the terms in \textit{stc}, and hence in \textit{ai4st}. 

Alongside this, RQ3 can also be answered positively. This small selection of unique research covers all potential purposes and levels of automation supported by AI. With regard to the types of AI, all but general AI (due to its non-existence) and evolutionary algorithms (which are currently not in focus) are represented. Additionally, 28~software testing targets are addressed, representing over 10\% of the extended \textit{stc} ontology.

In addition, \textit{ai4st} can be used as a research knowledge base. One straightforward application is elaborating on the classified research corpus: As the classification can be queried like a database, it is easy to formulate queries about the research corpus, such as which software testing targets are addressed or which research supports AI-assisted option automation, see the listings below.

\begin{lstlisting}[language=SPARQL, caption={All papers on AI-assisted options automation.}]
PREFIX rdf: <http://www.w3.org/1999/02/22-rdf-syntax-ns#>
PREFIX ai4st: <http://purl.org/ai4st/ontology#>
SELECT ?paper
WHERE {
  ?paper ai4st:hasLevel ai4st:AI-assisted_options .
}
\end{lstlisting}

\begin{lstlisting}[language=SPARQL, caption={All software testing targets addressed by research papers.}]
PREFIX rdf: <http://www.w3.org/1999/02/22-rdf-syntax-ns#>
PREFIX ai4st: <http://purl.org/ai4st/ontology#>
SELECT DISTINCT ?target
WHERE {
  ?paper rdf:type ai4st:ResearchPaper .
  ?paper ai4st:hasTarget ?target .
}
\end{lstlisting}

\section{Outlook}\label{Sec_Outlook}

This paper describes ongoing work on representing the exhaustive body of knowledge on software testing using an in-depth ontology. This ontology can also form the basis for exploring new research fields in software testing, such as the emerging area of using AI techniques and tools in software testing (ST). To this end, the paper presents the initial versions of  
\begin{itemize}
	\item the \textit{stc} ontology on software testing concepts, which is mainly based on the ISTQB Glossary and completed with SEVOCAB and proprietary software testing vocabulary.
	\item the \textit{ai4st} ontology that classifies AI for ST research according to the \textit{purpose} of the AI-based solution, the software testing \textit{target} being addressed, the \textit{type} of AI being used, and the \textit{level} of automation achieved. 
	\item an exemplary SLR on AI for ST, revealing 38 original research papers classified in the \textit{ai4st} ontology.
\end{itemize}

The research results including the ontologies \textit{stc} and \textit{ai4st} as well as the paper selections from the SLR are  available online for reuse and further uptake. The \textit{ai4st} can be used not only to understand the concepts in this research field better, but also to explore research related to a specific aspect, such as all papers on agentic AI for ST, using SPARQL queries.

The next step will be to extend \textit{stc} to cover the remaining ISTQB terms, to further refine \textit{ai4st} to cover more AI-related details, and and carefully revise new term and synonym candidates stemming from SLRs for potential addition. This will form the basis for classifying further research results on the application of \textit{ai4st} including \textit{stc}.

\subsubsection*{Acknowledgement} The ideas presented in this paper were developed through constructive dialogue within the Testing, Analysis and Verification (TAV) section of the German Informatics Society (GI), the German Testing Board (GTB), and the International Software Testing Qualifications Board (ISTQB). While the author wrote this paper independently, she acknowledges that the writing process was aided by DeepL to fine-tune the wording. The author has no competing interests to declare that are relevant to the content of this article.

\bibliographystyle{splncs04}
{\footnotesize
\bibliography{TestingAI_Schieferdecker}
}

\end{document}